\documentclass[namedreferences]{solarphysics}
\usepackage[optionalrh]{spr-sola-addons} 
\usepackage{graphicx}        
\usepackage{color}           
\usepackage{url}             

\usepackage[unicode=true, pdfusetitle, bookmarks=true, bookmarksnumbered=false, bookmarksopen=false, breaklinks=true, pdfborder={0 0 0}, backref=false, colorlinks=true]{hyperref}

\begin{document}

\begin{article}

\begin{opening}

\title{Formation of Coronal Shock Waves}

\author{S.~\surname{Luli\'c}$^{1}$\sep
        B.~\surname{Vr\v{s}nak}$^{2}$\sep
        T.~\surname{\v{Z}ic}$^{2}$\sep
        I.W.~\surname{Kienreich}$^{3}$\sep
        N.~\surname{Muhr}$^{3}$\sep
        M.~\surname{Temmer}$^{3}$\sep
        A.M.~\surname{Veronig}$^{3}$
       }
\runningauthor{S. Luli\'c {\it et al.}}
\runningtitle{Formation of Coronal Shocks}

   \institute{
$^{1}$ Karlovac University of Applied Sciences, Trg J.J. Strossmayera 9, HR-47000 Karlovac, Croatia
                     email: \url{slulic@vuka.hr}
                     \\
$^{2}$ Hvar Observatory, Faculty of Geodesy, University of Zagreb, Ka\v{c}i\'{c}eva 26, HR-10000 Zagreb, Croatia
                     email: \url{bvrsnak@geof.hr} email: \url{tzic@geof.hr}
                     \\
$^{3}$ IGAM, Institute of Physics, University of Graz, Universit{\"a}tsplatz 5, A-8010 Graz, Austria
                     email: \url{ines.kienreich@uni-graz.at} email: \url{nicole.muhr@edu.uni-graz.at} email: \url{manuela.temmer@uni-graz.at} email: \url{astrid.veronig@uni-graz.at}
                    \\
             }

\begin{abstract}
Numerical simulations of magnetosonic wave formation driven by an expanding cylindrical piston are performed to get better physical insight into the initiation and evolution of large-scale coronal waves caused by coronal eruptions. Several very basic initial configurations are employed to analyze
intrinsic
characteristics of the MHD wave formation that do not depend on specific properties of the environment. It turns out that these simple initial configurations result in piston/wave morphologies and kinematics that
reproduce common
characteristics of coronal waves. In the initial stage the wave and the expanding source-region cannot be clearly resolved, {\it i.e.} a certain time is needed before the wave detaches from the piston. Thereafter, it continues to travel as a so-called ``simple wave".
During the acceleration stage of the source-region inflation, the wave is driven by the piston expansion, so its amplitude and phase-speed increase, whereas the wavefront profile steepens.
At a given point, a discontinuity forms in the wavefront profile, {\it i.e.} the leading edge of the wave becomes shocked. The time/distance required for
the shock formation is shorter for a more impulsive source-region expansion. After the piston stops, the wave amplitude and phase-speed start decreasing. During the expansion, most of the source region becomes strongly rarified, which reproduces the coronal dimming left behind the eruption. On the other hand, the density increases at the source-region boundary, and stays enhanced even after the expansion stops, which might explain stationary brightenings that are sometimes observed at the edges of the erupted coronal structure. In addition, in the rear of the wave a weak density depletion develops, trailing the wave, which is sometimes observed as weak transient coronal dimming. Finally, we find a well defined relationship between the impulsiveness of the source-region expansion and the wave amplitude and phase speed.
The results for the cylindrical piston are also compared with the outcome for a planar wave that is formed by a one-dimensional piston, to find out how different geometries affect the evolution of the wave.
\end{abstract}
\end{opening}

\section{Introduction}
     \label{S-Introduction}

Explosive expansion of coronal structures associated with CME/flare eruptions frequently creates large-scale large-amplitude waves and shocks in the solar corona (for recent reviews, presenting detailed overview of various aspects of this phenomenon, see \opencite{warmuth07}; \opencite{vrs08cliver}; \opencite{wills09}, \opencite{warmuth10}; \opencite{gallagher11}; \opencite{zhukov11}; \opencite{patsourakos12}). These global disturbances are observed as EUV coronal waves, chromospheric Moreton waves, type II radio bursts, moving soft X-ray, and/or radio sources (see \opencite{warmuth04a}; \opencite{vrs06shock3nov}; \opencite{olmedo12}), as well as sharp fronts in the white-light coronagraphic CME images ({\it e.g.} \opencite{ontiveros09}). In recent years, this phenomenon was the subject of many studies that focused on various observational and theoretical aspects, including the morphology, kinematics, source-region characteristics, shock formation, three-dimensional propagation, {\it etc.} (for a brief overview of recent research activities see, {\it e.g.}, Section 8 in \opencite{IAU09} and Section 9 in \opencite{IAU12}).

At low coronal heights, where disturbances are observed in the EUV range, waves usually become recognizable at a distance of $\approx$\,100\,--\,200 Mm from the source active region ({\it e.g.} \opencite{veronig08}; \opencite{patsourakos09}; \opencite{ines11}; \opencite{muhr11}). Thus, EUV waves are observed while propagating through a quiet corona, where the magnetic field is predominantly vertical. Consequently, a low-coronal wave segment can be considered as a perpendicular magnetohydrodynamical (MHD) wave (magnetosonic wave). Typical velocities of EUV waves are a few hundred km\,s$^{-1}$ (for details see, {\it e.g.}, \opencite{thompson09}; \opencite{warmuth11} and references therein).

New detailed observations reveal that the wave amplitude initially increases and at the same time the wave accelerates. Eventually, after a phase of approximately constant speed, the wave decelerates to velocities typically around 200\,--\,300 km\,s$^{-1}$ ({\it e.g.} \opencite{long08}; \opencite{muhr12}; \opencite{temmer12}), where faster waves show a stronger deceleration \cite{liu10,kozarev11,ma11,warmuth11,cheng12,olmedo12}. It was also found that waves of  higher speed have larger amplitude \cite{ines11}. During the constant-speed and deceleration stage, the amplitude of the perturbation decreases whereas its profile broadens. Such behavior is usually interpreted as a typical signature of a freely propagating ``simple wave" (for terminology we refer to \opencite{vrs05EOS}; \opencite{warmuth07}).

The fastest waves are frequently accompanied by type II radio bursts ({\it e.g.} \opencite{klassen00}; \opencite{biesecker02}; \opencite{warmuth04b}; \opencite{veronig06}; \opencite{vrs06shock3nov}; \opencite{muhr10}; \opencite{ma11}; \opencite{kozarev11}), which reveal the formation of a coronal MHD shock. Such waves may also generate Moreton waves ({\it e.g.} \opencite{warmuth04a}; \opencite{vrs06shock3nov}; \opencite{muhr10}; \opencite{asai12}; \opencite{shen12}) if the pressure jump at the shock front is strong enough to push the inert chromospheric plasma downwards, {\it i.e.} if the shock amplitude is high enough.

Generally, coronal waves and shocks could be generated by the source-region expansion either related to a coronal mass ejection (CME), or a pressure pulse caused by the flare-energy release (for a discussion see \opencite{vrs08cliver}). Whereas in many events the source-region expansion could be clearly identified with the impulsive-acceleration stage of a CME ({\it e.g.} \opencite{patsourakos10}; \opencite{veronig10}; \opencite{grechnev11}; \opencite{kozarev11}); in some cases there are indications that the shock is initiated by a flare ({\it e.g.} \opencite{vrs06shock3nov}; \opencite{magdalenic10}; \opencite{magdalenic12}).

Whatsoever the driver is, perpendicular MHD shocks are created by plasma motion perpendicular to the magnetic field. For example, a supersonic motion of small-scale ejecta would produce a shock (see, {\it e.g.}, \opencite{klein99}), in a similar manner as that in which supersonic projectiles create shocks in the air. However, in the solar corona a much more suitable process is a source-region expansion which acts as the three-dimensional (3D) piston. If the expansion is impulsive enough it creates a large-amplitude perturbation, whose leading edge steepens due to non-linear effects, {\it i.e.} wave elements of higher amplitude move faster. Eventually, a discontinuity occurs in the wavefront profile, meaning that the shock is formed. Whereas the 1D MHD piston problem (planar wave) can be solved analytically \cite{mann95,V&L00a}, an analogous 2D or 3D problem can be treated analytically only by applying severe assumptions and approximations, and after all, a numerical evaluation is needed in any case (see, {\it e.g.}, \opencite{zic08}; \opencite{afanasyev11}).

Thus, numerical MHD simulations are required to study a 2D and 3D piston mechanism for the magnetosonic-wave generation.
Bearing in mind that the wave formation and evolution are strongly influenced, or even dominated, by physical properties of the environment and the characteristics of the driver itself, there are two alternatives in approaching this complex physical problem. One way is to set up the initial conditions as closely as possible to the real situation in which a particular wave has occurred, and to perform a full 3D simulation that provides a detailed quantitative analysis of the specific event. Such an approach, providing detailed insight into the physics behind a particular event, including coronal diagnostics, were performed by, {\it e.g.}, \inlinecite{uchida73}, \inlinecite{wang00}, \inlinecite{wu01}, \inlinecite{ofman02}, \inlinecite{ofman07}, \inlinecite{cohen09}, \inlinecite{schmidt10}, \inlinecite{downs11}, \inlinecite{selwa12}. Another way is to start from a somewhat simplified initial situation, which provides more extensive parametric studies and gives a more general view on the problem. In this type of simulation the CME is usually represented by an erupting 2D structure, anchored in the inert photosphere (see, {\it e.g.}, \opencite{chen02}, \opencite{chen05}, \opencite{pomoell08}, \opencite{wang09}).

In this article we consider some simple initial configurations/geometries to get an insight into the most basic characteristics of nonlinear processes governing
the MHD wave formation and evolution in general. The idea is to isolate the basic processes that stand behind the wave formation in an idealized surrounding, {\it i.e.} to identify effects that are present regardless of the specific properties of the environment. In the follow-up article, more realistic configurations will be considered, including the chromosphere/corona density and Alfv\'en speed profile, magnetic field line-tying, and the arcade expansion accompanied by an upward motion. These more advanced simulations will be compared with the results presented in this article, which will help us to distinguish the effects that are intrinsic to the MHD wave formation from those governed by the environment.

Special attention is paid to the wavefront steepening, {\it i.e.}, the shock formation process, in a planar and cylindrical geometry. For the simulations we employ the Versatile Advection Code (VAC:  \opencite{toth96}; \opencite{Goedbloed03}). This numerical code was developed at the Astronomical Institute at Utrecht, in collaboration with the FOM Institute for Plasma Physics, the Mathematics Department at Utrecht, and the Centrum Wiskunde and Informatica (CWI) at Amsterdam. It is suitable for the analysis of a broad spectrum of astrophysical phenomena, including magnetohydrodynamic (MHD) shock waves.

In Section 2 the model employed and the simulation procedure are briefly described. In Section 3 we present the results, first considering a planar geometry, so that the outcome can be compared with the analytical results, and then switching to a cylindrical geometry, which is more closely related to a coronal-arcade eruption or a coronal-loop expansion. In Section 4 we discuss the results and compare them with observations.

\section{The Model} 
      \label{S-model}

In the following, we consider perpendicular magnetosonic waves, where we focus on a planar and cylindrical geometry. This allows us to set the magnetic field in the $z$-direction, whereas the $x$ and $y$ magnetic-field components, as well as the $z$-component of the velocity, are always kept zero ($B_x=0$, $B_y=0$, $v_z=0$). Furthermore, all quantities are invariant along the $z$-coordinate, {\it i.e.} we perform 2.5D simulations, where the input and the basic output quantities are the density [$\rho$] the momentum [$m_{x}=\rho v_{x}$, $m_{y}=\rho v_{y}$] and the magnetic field [$B_z$]. Note that although we perform 2.5D simulations, physically it is a one-dimensional problem.

We use a two-dimensional [2D] numerical mesh containing $995\times995$ cells, supplemented by two ghost-cell layers at each boundary, which are used to regulate the boundary conditions (thus, the complete grid consists of $999\times999$ cells).
We apply continuous boundary conditions, meaning that gradients of all quantities are kept zero by copying the variable values from the edge of the mesh into the ghost cells. All quantities are normalized, so that distances are expressed in units of the numerical-box length [$L=1$], velocities are normalized to the Alfv\'en speed [$v_A$], and time is expressed in terms of the Alfv\'en travel time over the numerical-box length [$t_A=L/v_A$]. We apply the approximation $\beta=0$, where $\beta$ is the plasma-to-magnetic pressure ratio. The origin of the coordinate system is set at the numerical-box center.

We will consider two basic initial configurations, resulting in a planar wave and a cylindrical one. In the planar option, all quantities are invariant in the $y$-direction, {\it i.e.} quantities depend only on the $x$-coordinate. In the cylindrical option, all quantities depend only on the $r$-coordinate, where $r^2=x^2+y^2$.

In all runs, we set up the simulation as an initial-value problem, starting from an unstable magnetic-field configuration of the source region. Thus, the source-region expansion is not fully under control, {\it i.e.} we do not prescribe the time-profile of the ``driver" motion. The overall characteristics of the source-region expansion (the acceleration impulsiveness and the maximum speed) are regulated only indirectly, by increasing or decreasing the initial force imbalance. More precisely, we start from an initial configuration where the force balance is distressed by the excess magnetic pressure, {\it i.e.} the space--time evolution of the plasma flow is entirely determined by the initial spatial profile of $B^2$.

Specifically, we set a ``parabolic" profile for the initial magnetic field within the source region:
\begin{equation}
B_z(x)=\sqrt{B_{0}^{2}-b\,x^{2}}\,,
\end{equation}
where $B_0$ represents the magnetic field at $x=0$ and $b(x)$ defines the field strength profile within the source region. We employ the form $b=(B_0^2-B_e^2)/x_0^2$, where $x_0$ is the initial source-region size and $B_e$ represents the external magnetic field strength outside the source region. In the cylindrical configuration we use the same function, only replacing $x$ by $r$. The initial magnetic-field profile is drawn in Figure~\ref{1Dprofiles}a by a red line. For the initial source-region size we take $x_0=0.1$; beyond $x=x_0$ we set $B_e=1$ and $\rho_e=1$. To make the source region more inert, and to better visualize the source region, we increased the density within the source region to $\rho=2$ (see the red line in Figure~\ref{1Dprofiles}c).

At the beginning, the plasma is at rest, $v=0$ (see the red line in Figure~\ref{1Dprofiles}e). The considered profile of $B_z$ is characterized by a magnetic-pressure gradient [$\partial(B_z^2/2\mu_0)/\partial x$] which causes the initial outward acceleration. The acceleration increases linearly from 0 at $x=0$ to a maximum value at the source-region boundary (hereinafter denoted also as ``contact surface", or ``piston"). The motion of the source-region boundary is tracked by identifying a surface within which the mass content equals the initial one.

In the cylindrical geometry we also use another initial magnetic-field profile for the source region. It has the form:
\begin{equation}
B_{z}(r)=
B_{0}\cos^{2}\left(\frac{\pi}{2}\frac{r}{r_{0}}\right)+B_{e},
\end{equation}
where $r_0$ represents the source-region size, and $B_{e}$ is the magnetic-field strength for $r>r_0$. In this case the magnetic pressure gradient and the initial acceleration are zero at the source-region center and at the source surface, whereas the peak value is attained within the source-region body, at $r/r_0=\pi/4$. The density is again set to $\rho=2$ within $r<r_0$, and $\rho=1$ for $r>r_0$.

Although we do not intend to reproduce directly any specific coronal structure, note that the hereafter presented evolution of the two  cylindrical configurations employed can depict, to a certain degree, the coronal-wave formation caused by a lateral expansion of the source region placed in a vertical magnetic field of a quiet corona. Such an expansion can occur, {\it e.g.} in the impulsive-acceleration stage of CMEs (the so-called ``lateral overexpansion"; see, {\it e.g.}, \opencite{kozarev11}; \opencite{patsourakos10}), or presumably, in legs of impulsively heated flaring loops.

\section{Results} 
      \label{S-result}

\subsection{Formation and Propagation of a Planar Shock }
      \label{S-1D}

   \begin{figure}    
 \centerline{
 \includegraphics[width=0.9\textwidth]{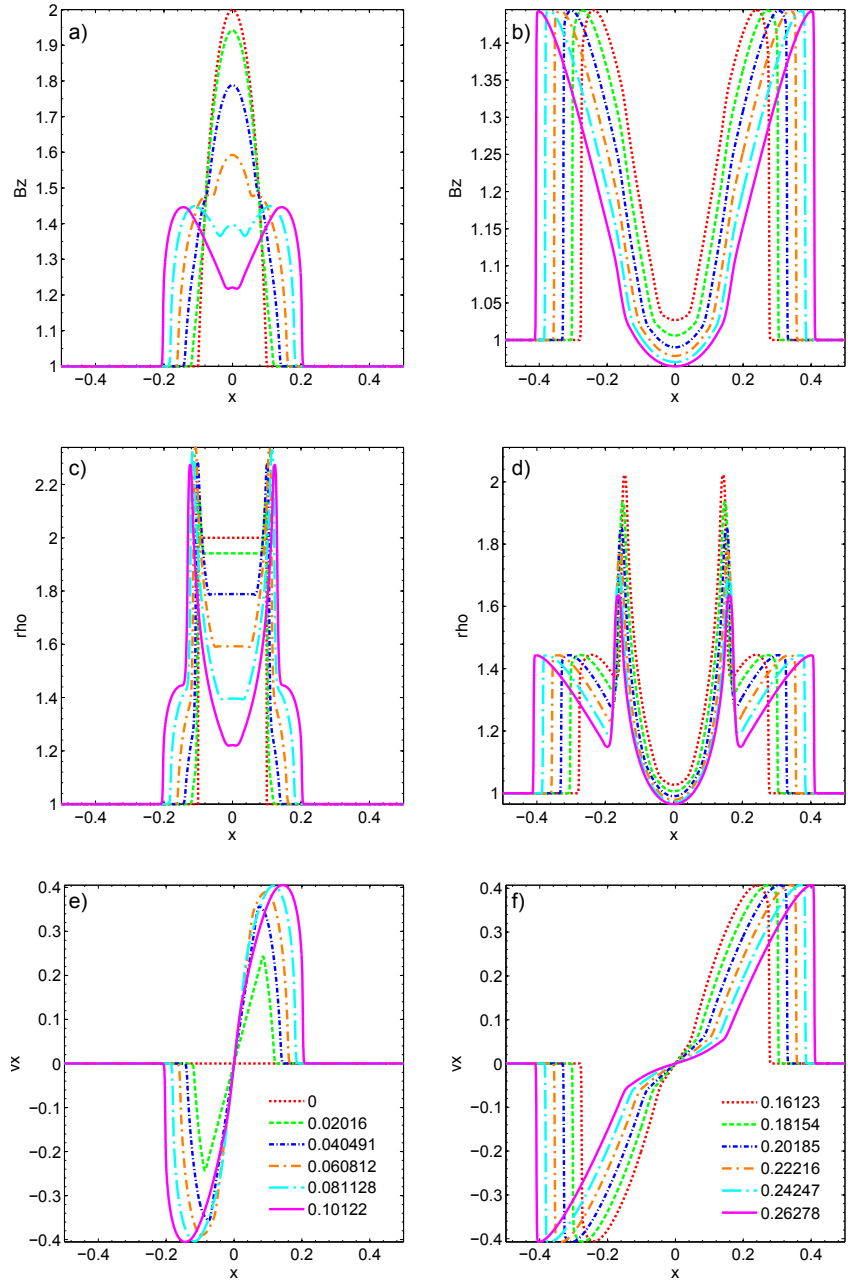}
              }
\caption{Formation and propagation of a perpendicular shock in the planar geometry: spatial profiles of the magnetic field (a, b), density (c, d), and flow speed (e, f). Left panels (a, c, e): the beginning of the wave formation (increasing wave amplitude); right panels (b, d, f): the shock formation phase (steepening of the wavefront profile). The initial magnetic field in the source-region center is $B_0=2$. Times are displayed in the inset. All quantities presented are normalized: distance $x$ is expressed in units of the numerical-box length [$L=1$], velocity $v_x$ is normalized to the Alfv\'en speed [$v_A$], and time $t$ is expressed in terms of the Alfv\'en travel time over the numerical-box length [$t_A=L/v_A$].
        }
 \label{1Dprofiles}
   \end{figure}

First we analyze the formation and propagation of a perpendicular shock in the planar geometry. The aim is to compare the numerical results with the analytic theory for planar MHD waves (\opencite{V&L00a}; \citeyear{V&L00b}) and to have the reference-results when the influence of the geometry on the results will be considered.

\begin{figure}    
 \centerline{
 \includegraphics[width=.6 \textwidth]{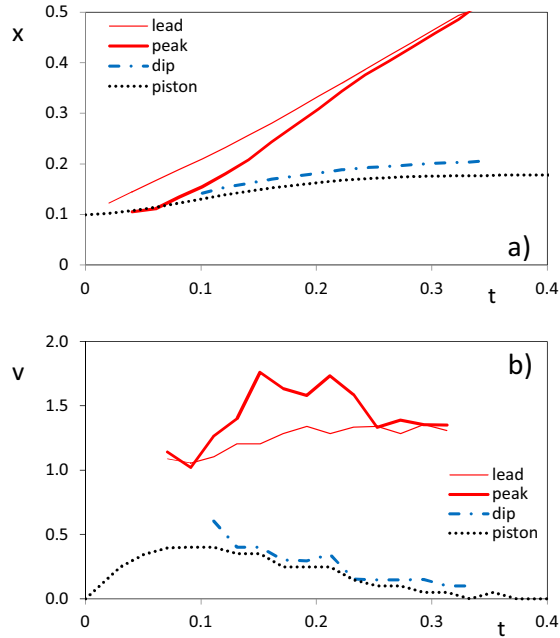}
              }
   \caption{ Kinematics of various wave features and the source-region boundary (thin-solid line -- the wavefront leading edge; thick-solid line -- the wave crest; dot-dashed line -- a trailing density dip; dotted line -- the source-region boundary): a) distance {\it versus} time; b) velocity {\it versus} time.
               }
 \label{1d kinem}
   \end{figure}

The formation and propagation of the wave is presented in Figure~\ref{1Dprofiles}. In Figures~\ref{1Dprofiles}a and b we show the magnetic-field profiles [$B_z(x)$] in Figures~\ref{1Dprofiles}c and d the density profiles [$\rho(x)$] and in Figures~\ref{1Dprofiles}e and f the flow-speed profiles [$v_x(x)$]. The graphs in the left column show the formation phase of the wave, whereas those on the right side represent the propagation phase. The kinematics of various features recognized in the density profiles in Figures~\ref{1Dprofiles}c and d (the wavefront leading edge, the wave peak, a dip between the wave and the piston, and the source-region boundary) are displayed in Figure~\ref{1d kinem}.

Due to the magnetic pressure gradient of the unstable initial configuration,
the source-region expansion starts immediately at $t=0$. The acceleration is strongest at the source-region surface, whereas the source-region center ($x=0$) stays at rest. Over most of the source-region volume, the density decreases due to the expansion, whereas close to the contact surface it increases due to the velocity gradient. The kinematics of this density peak closely follows the kinematics of the contact surface, just slightly lagging behind it. A peak density, $\rho=2.34$, is attained around $t=0.06$. The source-region boundary accelerates until $t\approx0.1$, attaining a speed of $v=0.4$. After that, it gradually slows down, and stops around $t\approx0.35$ (see kinematics shown in Figure~\ref{1d kinem}a). During the accelerated-expansion phase, the flow speed increases, attaining a value of $v\approx0.4$ around $t\approx0.1$ (Figure~\ref{1Dprofiles}e), {\it i.e.} the fastest flow elements are adjusted to the piston motion.

Ahead of the contact surface, a wavefront forms as a result of the source-region expansion. It can be easily recognized in the magnetic-field and density profiles shown in Figures~\ref{1Dprofiles}a and c. The wave detaches from the source region after $t\approx0.1$ ({\it i.e.} after the piston acceleration-phase ends), and continues to evolve as a freely propagating simple wave (for a hydrodynamic analog see Sections 101 and 102 in \opencite{L&L87}). Note that a dip in the density profile, formed between the wave peak and the contact-surface peak, never gets values $\rho<1$. On the other hand, the density in central parts of the source region becomes strongly depleted.

The wavefront steepens in time, whereas its amplitude remains constant, staying at values of $\rho=1.44$, $B=1.44$, and $v=0.42$, respectively. The shock formation is completed at $t\approx0.26$.

The kinematics of the wave leading edge, the wave peak, a rarefaction dip,
and the piston, measured from the density profiles shown in Figure~\ref{1Dprofiles}, are displayed in Figure~\ref{1d kinem}, revealing that the piston accelerates until $t\approx0.08$. Thereafter, it continues to move at an approximately constant speed of $v\approx0.4$ until $t\approx0.13$. During this period the wave amplitude increases (see Figure~\ref{1Dprofiles}) and the wave-crest phase speed increases from $w\approx1$ to $w=1.76$. At the same time, the wavefront leading edge moves at $w\approx1$. The wave crest reaches the leading edge, {\it i.e.} the shock formation is completed, around $t\approx0.25$. After that, the shock front moves at a speed of $w=1.35$, consistent with the Rankine--Hugoniot jump relations. The described evolution of the source/wave system and its kinematics is fully consistent with the analytical model presented by \inlinecite{V&L00a}.

After $t\approx0.13$, the source-region expansion gradually decelerates, and practically stops at $t\approx0.35$. A density dip between the wave peak and the piston, which forms around $t\approx0.11$, closely follows the kinematics of the source-region boundary, being only slightly faster then the piston.

   \begin{figure}    
 \centerline{
 \includegraphics[width=.6\textwidth]{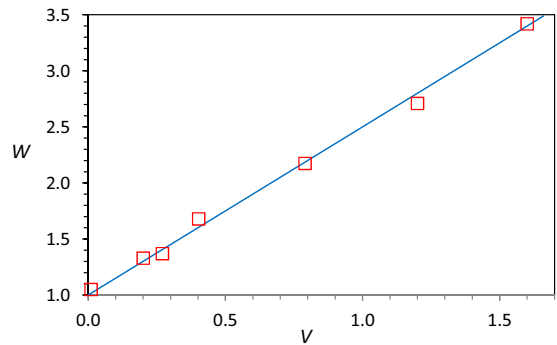}
              }
   \caption{ Relationship between the wave speed [$w$] and the flow speed [$v$] for the planar wave. Numerical results (red squares) are compared with the analytical relationship $w=1+3v/2$ (blue line) derived by Vr\v snak and Luli\'c (2000).
               }
 \label{1d}
   \end{figure}

We repeated the simulations using various values of $B_0$, to analyze how the evolution of the piston/wave system depends on the impulsiveness of the source-region expansion. A stronger $B_0$ results in a more impulsive source-region acceleration, which leads to a higher shock amplitude and Mach number. Furthermore, the shock is created earlier and closer to the piston, so in the case of extremely impulsive expansions, the shock-sheath region and the source region cannot be clearly resolved.

In Figure~\ref{1d} we show the dependence of the phase speed [$w$] of the perturbation-segment at the wave crest (before being shocked) as a function of the corresponding flow speed $v$. In the graph we display the results for $B_0=1.5$, 2.0, 3.0, and 5.0. For $B_0=2$ we also measured $w$ and $v$ at several suitable wavefront-segments ahead of the wave crest (the lowest $v$-values in Figure~\ref{1d}). The results are fully consistent with the outcome of the analytical theory for $\beta=0$ presented by \inlinecite{V&L00a}, where the relationship $w=1+3v/2$ was established.

Whereas in the non-shocked phase the wave behavior is consistent with the analytical theory, in the shocked phase for $B_0\gtrsim3$, corresponding to $w\gtrsim1.8$, the numerical results start to deviate from the analytical Rankine--Hugoniot relations, the disagreement increasing with the increasing amplitude. The equation of continuity and the relationship between Mach number and the downstream flow speed behave as expected, but the relationship between Mach number and the downstream/upstream density jump deviates from the analytical results. This is probably due to numerical effects and the fact that at very high values of $B_0$ it becomes impossible to clearly resolve the compression at the source-region surface and the shock itself.

\subsection{Cylindrical Geometry} 
      \label{S-cyl}

\subsubsection{Wave Formation} 
      \label{S-form}

   \begin{figure}    
 \centerline{
 \includegraphics[width=0.9\textwidth]{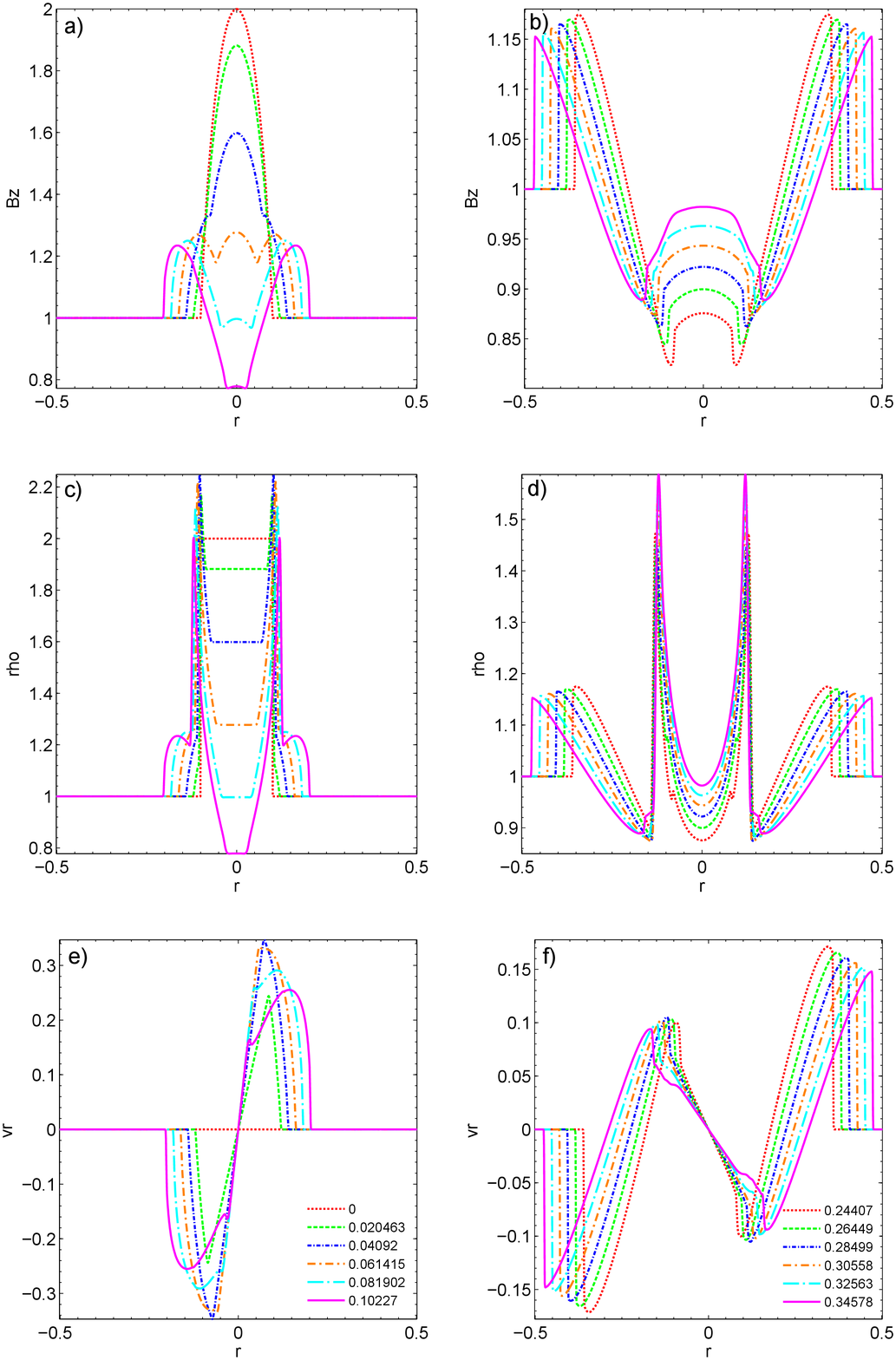}
              }
\caption{Formation and propagation of the perpendicular shock in the cylindrical geometry for the initial magnetic field profile given by Equation (1) with $B_0=2$: spatial profiles of the magnetic field (a, b), the density (c, d), and the flow speed (e, f). Left panels (a, c, and e) show the beginning of the wave formation (increasing wave amplitude); right panels (b, d, and f) present the shock formation phase (steepening of the wavefront profile). Times are displayed in the inset. All quantities presented are normalized: distance $r$ is expressed in units of the numerical-box length [$L=1$], velocity $v_r$ is normalized to the Alfv\'en speed [$v_A$], and time $t$ is expressed in terms of the Alfv\'en travel time over the numerical-box length [$t_A=L/v_A$].
        }
   \label{cyl_profiles}
   \end{figure}

In Figure~\ref{cyl_profiles} the formation and propagation of the wave in the cylindrical geometry is presented. Spatial profiles of the magnetic field [$B_z(r)$] are shown in Figures~\ref{cyl_profiles}a and b, the density profiles [$\rho(r)$] are presented in Figures~\ref{cyl_profiles}c and d, whereas in Figures~\ref{cyl_profiles}e and f the flow speed [$v_r(r)$] is displayed. The initial magnetic-field and density profiles are defined in the same way as in the planar case (Equation (1)), only replacing $x\rightarrow r$. In the left column the wave formation phase is shown, whereas the right column represents the propagation phase. In Figure~\ref{cyl_kinem}, the kinematics of various features recognized in Figure~\ref{cyl_profiles} are shown.

As in the planar case, the source-region expansion starts immediately at $t=0$, the acceleration being strongest at the source-region surface. The source-region center ($r=0$) remains at rest at all times (Figures~\ref{cyl_profiles}e and f). The density within the source region starts to decrease due to the expansion, whereas at the contact surface it increases due to the flow-speed gradient. The source-region expansion initially accelerates, attaining $v=0.28$ at $t\approx0.07$ (see Figure~\ref{cyl_kinem}). Note that the acceleration phase is shorter than in the planar case, the peak speed is considerably lower, and the $v$\,$\approx$\,const. phase is absent. After attaining the maximum speed, the piston gradually slows down, stops around $t\approx0.2$, and then retreats slowly towards the initial position (see the kinematics shown in Figure~\ref{cyl_kinem}). During the accelerated-expansion phase, the flow speed increases, reaching $v=0.28$ around $t\approx0.04$ (Figure~\ref{cyl_profiles}e).
Note that, unlike in the planar geometry, the plasma flow is not fully synchronized with the piston motion.

Ahead of the contact surface, the wavefront forms as a result of the source-region expansion. It can be readily recognized in the magnetic field and density profiles shown in Figures~\ref{cyl_profiles}a and c. The wave detaches from the source region around $t\approx0.08$, having an amplitude in $\rho$ and $B$ of around 1.22. After that, the perturbation continues to propagate as a freely propagating simple wave, but unlike in the planar case, the amplitude of the wave decreases with distance (Figure~\ref{cyl_profiles}). The wavefront steepens with time, whereas the peak flow-speed decreases. A discontinuity in the leading-edge profile occurs, {\it i.e.} the shock formation begins at $t\approx0.15$. The shock is fully completed at $t\approx0.28$, when it has an amplitude of $\rho=1.16$ and $v=0.16$.

Note that flows within the source region are more complex than in the planar configuration. We also stress that a dip in the density profile, which forms between the wave peak and the contact-surface, now deepens to a value of $\rho=0.88$, {\it i.e.} the rarefaction region forms ($\rho<1$), as in the case of cylindrical hydrodynamic waves (see Section 102 in \opencite{L&L87} and references therein).

\subsubsection{Wave Kinematics} 
      \label{S-kinem}

   \begin{figure}    
 \centerline{
 \includegraphics[width=.6 \textwidth]{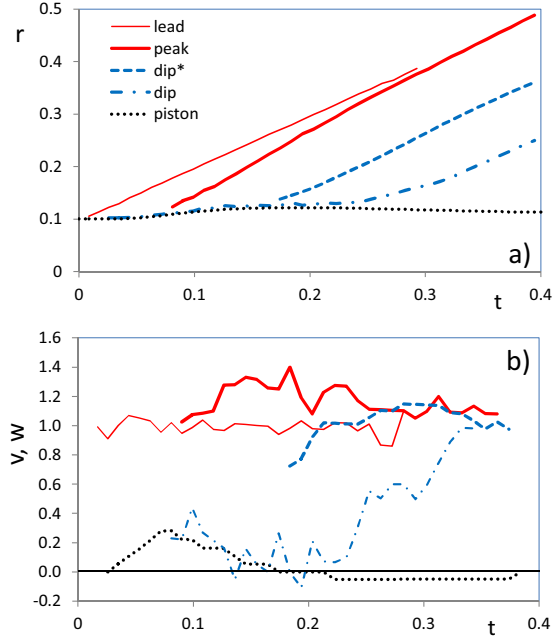}
              }
   \caption{ Kinematics of the wave and the source-region boundary derived from the density profiles shown in Figure~\ref{cyl_profiles}: a) distance {\it versus} time; b) velocity {\it versus} time. Thin-solid line -- the wavefront leading edge; thick-solid line -- the wave peak; dashed line -- a deep front measured at $\rho=1$; dot-dashed line -- the dip minimum; dotted line -- the source-region boundary. All quantities presented are normalized: radial distance $r$ is expressed in units of the numerical-box length [$L=1$], velocities $v$ and $w$ are normalized to the Alfv\'en speed [$v_A$], and time $t$ is expressed in terms of the Alfv\'en travel time over the numerical-box length [$t_A=L/v_A$].
               }
 \label{cyl_kinem}
   \end{figure}

The  kinematics of the piston and the wave, estimated from the density profiles displayed in Figure~\ref{cyl_profiles}, are shown in Figure~\ref{cyl_kinem}. Comparing Figures~\ref{cyl_profiles} and ~\ref{cyl_kinem} one finds that during the piston acceleration the wave amplitude first increases, but then starts to decrease, even before the piston reaches its maximum velocity. The phase speed of the wave crest increases from $w\approx1$ to $w\approx1.3$ at $t\gtrsim0.15$, thereafter it gradually decreases.

A dip between the wave peak and the piston, which forms around $t\approx0.08$, first closely follows the piston kinematics, but then, after the dip becomes characterized by $\rho<1$ at $t\approx0.2$, it ``detaches" from the piston and attains a speed of $w\approx1$ around $t\approx0.35$. Note that a segment of the dip characterized by $\rho=1$ moves at a speed of $w\approx1$ all the time.

   \begin{figure}    
 \centerline{
 \includegraphics[width=.6 \textwidth]{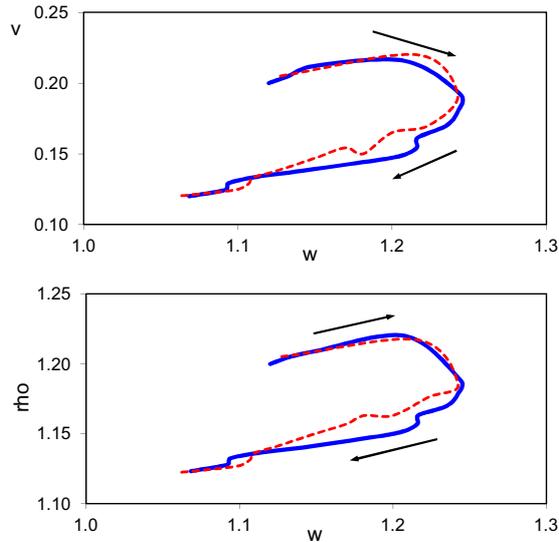}
              }
   \caption{ The evolution of:
   a) downstream flow-speed amplitude {\it versus} phase-speed of the wave crest;
   b) density amplitude {\it versus} phase-speed of the wave crest. Solid-blue and dashed-red lines show
    results for the initial configuration defined by Equations (1) and (2), respectively, with $B_0=2$. Arrows indicate the course of the temporal sequence.
    Velocities $v$ and $w$ are normalized to the Alfv\'en speed [$v_A$].
               }
 \label{hyster}
   \end{figure}

The relationship between the shock speed and the downstream flow speed in the cylindrical geometry is more complex than in the planar case. This is illustrated in Figure~\ref{hyster}a, where the downstream flow-speed [$v$] is shown {\it versus} the phase-speed of the wave crest [$w$]. Analogously, in Figure~\ref{hyster}b we present the dependence of the downstream peak density [$\rho$] on the phase speed [$w$]. Note that the displayed values are based on smoothed curves $w(t)$, $v(t)$, and $\rho(t)$. The presented graphs show that initially, during the wave formation phase, the wave phase-speed increases, while the amplitudes of $v$ and $\rho$ are almost constant, showing only a slight increase. Then, the wave speed [$w$] remains almost constant, whereas the wave amplitude decreases. Eventually, in the third step, both the wave propagation speed and its amplitude decrease.

The highest values of $v$ and $\rho$ are attained roughly at a time when the maximum speed of the piston is reached. The shock formation starts, {\it i.e.} a discontinuity occurs at the leading edge of the wave profile, around the ``nose" of the curves presented in Figure~\ref{hyster}, which also approximately coincides with the end of the piston expansion. Thus, roughly speaking, the upper branch of the $v(w)$ and $\rho(w)$ curves corresponds to the ``driven phase" of the wave, whereas the lower branch represents the ``decay" of a freely-propagating simple wave. The shock is completed when the wave speed attains a value of $w\approx1.11$ at the lower branch of the curve.

\subsubsection{Piston Impulsiveness} 
      \label{S-impuls}

   \begin{figure}    
 \centerline{
 \includegraphics[width=.6 \textwidth]{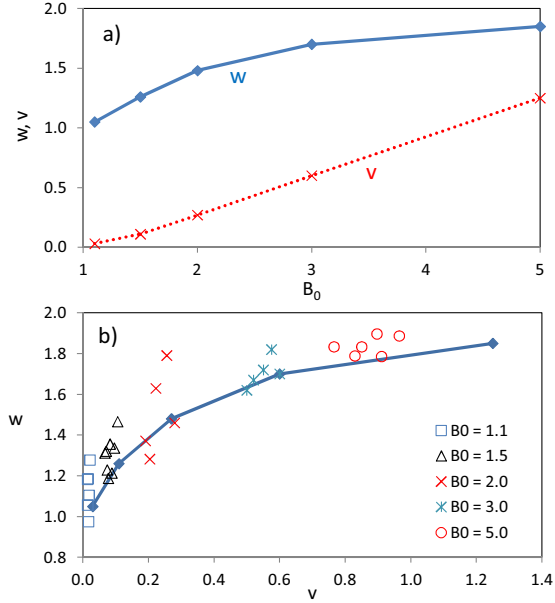}
              }
   \caption{ a) Peak phase-speed [$w$] of the wavefront (solid line and diamond symbols) and the speed [$v$] of the piston (dotted line and crosses) for five values of $B_0$. b) The phase-speed of the wavefront around the peak amplitude (squares, triangles, crosses, asterisks, and diamonds; the values of $B_0$ are given in the inset) presented as a function of corresponding flow speed. The solid line represents the peak velocity of the wave as a function of the maximum speed of the piston for the same five values of $B_0$. Velocities $v$ and $w$ are normalized to the Alfv\'en speed [$v_A$].
               }
 \label{w(v)}
   \end{figure}

We repeated the procedure presented in Sections \ref{S-form} and \ref{S-kinem} for several values of the maximum magnetic-field strength in the source-region center [$B_0$] to inspect the role of the impulsiveness of the piston expansion. In particular, we applied $B_0=1.1$, 1.5, 2.0, 3.0, and 5.0. A stronger field $B_0$ causes a more impulsive acceleration of the piston, which results also in a higher wave amplitude and wave-crest speed, and consequently, an earlier formation of the shock. On the other hand, the evolution of the system, as well as the relationship between different parameters, does not depend qualitatively on the impulsiveness of the piston acceleration. Morphologically, the main difference between very impulsive piston accelerations and more gradual ones is that in the former case the shock forms very close to the contact surface. Because of this, in the case of a very impulsive source-region expansion, it is not possible to follow the shock-formation phase, since the wavefront and the piston cannot be resolved. On the other hand, we note that for $B_0=1.1$ and 1.5 the shock did not form within the numerical box, which implies that in reality, particularly bearing in mind dissipative effects, the coronal shock would not be formed if the source-region acceleration is not impulsive enough.

In Figure~\ref{w(v)}a the peak velocity of the wavefront is compared with the peak velocity of the piston for different values of $B_0$. In the considered range, the wave speed is much larger than the piston speed, so the distance from the wavefront and the piston rapidly increases. However, the graph shows that beyond $B_0\approx1.5$ maximal piston velocities are proportional to $B_0$, whereas the wave speed shows a nonlinear trend, {\it i.e.} the slope of the $w(B_0)$ curve gradually decreases. This implies that for a very impulsive expansion of the source region one can expect that the velocities of the shock and the piston become comparable, and that the separation is small.

In Figure~\ref{w(v)}b we present the dependence of the maximum speed of the wave crest on the maximum speed of the corresponding downstream flow speed for all considered values of $B_0$. The displayed data points are numerical values from the ``nose" of the $w(v)$ curves (non-smoothed), analogous to the one shown in Figure~\ref{hyster}a for $B_0=2$. Peak values of $w$, estimated from smoothed $w(v)$ curves, are presented as a function of the maximum speed of the piston by a solid line. Note that this ``piston-curve" is shifted to the right with respect to the presented data points, implying that the piston speed is somewhat higher than the flow speed. The difference increases with the increasing piston speed, {\it i.e.} with the impulsiveness of the  expansion.

The main feature of Figure~\ref{w(v)}b is that the relationship between the wave speed and the downstream flow speed [$w=1+3v/2$] is not valid in cylindrical geometry. The $w(v)$ dependence is not linear, but is closer to a power-law. A least-square fit of the form $w-1=av^b$ gives $w=1+0.9\,v^{0.45}$, with a correlation coefficient of $R=0.91$. On the other hand, the relationship between the maximum wave speed and the maximum piston speed is well described ($R=0.99$) by the function $w=1+1.26\,v^{1/3}$.

\subsubsection{Initial Configuration} 
      \label{S-trigon}

   \begin{figure}    
 \centerline{
 \includegraphics[width=0.9\textwidth]{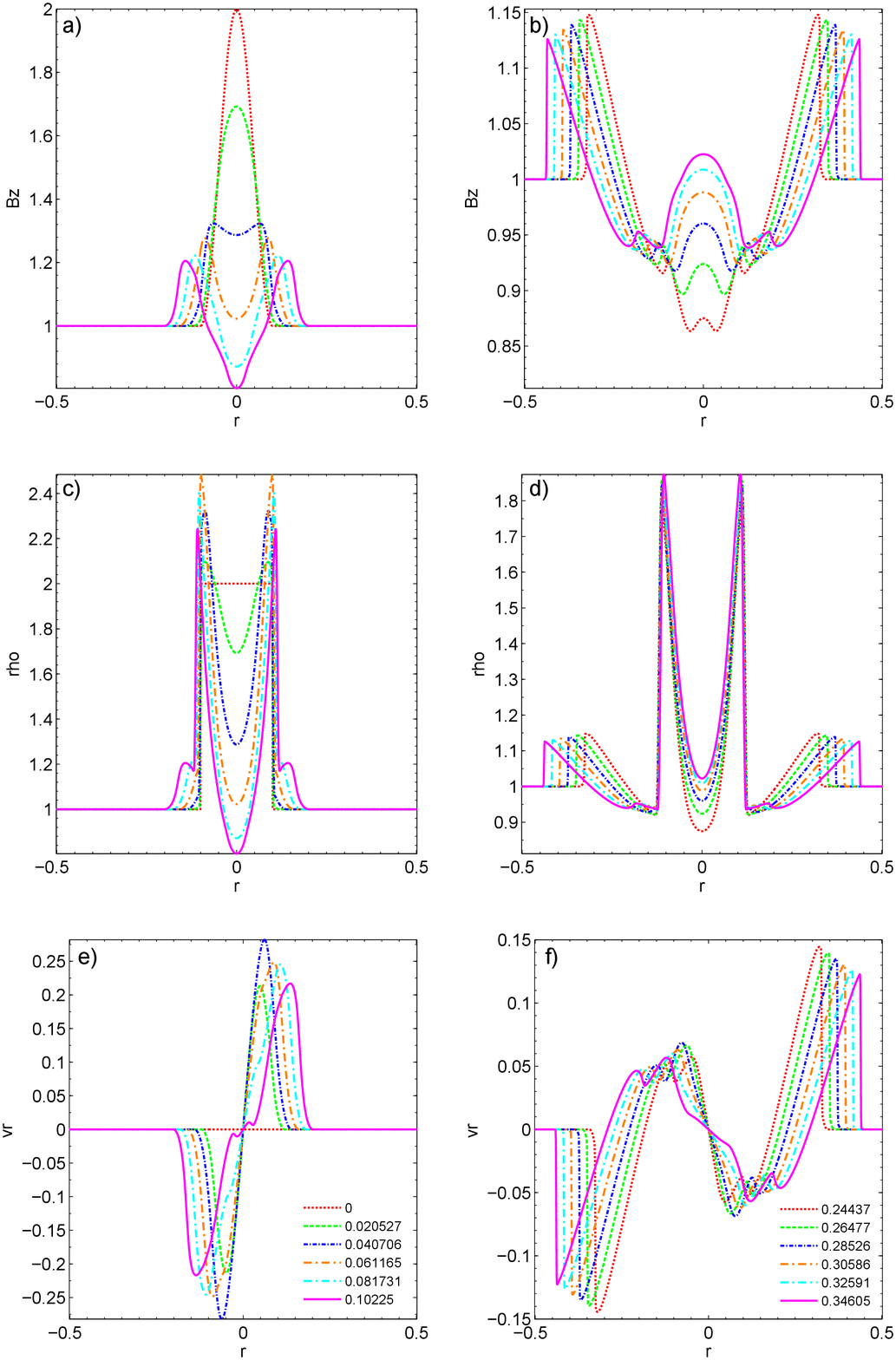}
              }
\caption{Formation and propagation of the perpendicular shock in the cylindrical geometry for the initial magnetic-field profile given by Equation (2), with $B_0=2$: spatial profiles of the magnetic field (a, b), density (c, d), and flow speed (e, f). Left panels (a, c, and e) show the beginning of the wave formation (increasing wave amplitude); right panels (b, d, and f) present the shock formation phase (steepening of the wavefront profile). Normalized times are displayed in the inset.
        }
   \label{trig_profiles}
   \end{figure}

To check how much the initial magnetic-field structure in the source region affects the process of the wave formation and evolution, we also applied the magnetic-field configuration defined by Equation (2). In this case, the steepest magnetic-pressure gradient is located within the source region, {\it i.e.} not at its edge as was the case described by Equation (1). The outcome of the simulation for $B_0=2$ is presented in Figure~\ref{trig_profiles}.

The analysis of the data in Figure~\ref{trig_profiles} shows that there are no significant differences in the overall wave kinematics for the two configurations considered (thus, the corresponding graphs are not shown). To illustrate the similarity of the two kinematics, we also included in Figure~\ref{hyster} the results concerning the wave formation/propagation resulting from the configuration defined by Equation (2).

However, Figure~\ref{trig_profiles} reveals a considerable difference in the morphology of the evolving piston/wave system. The strongest magnetic pressure gradient is initially located within the source region, whereas it is zero at its edge. Thus, the highest acceleration occurs at $r<r_0$, and consequently, the initial compression forms within the source region. This causes a more complex flow pattern within the source and obscures the initiation of the wave, {\it i.e.} the source-region boundary and the wave cannot be clearly distinguished, and the term piston becomes meaningless.

The wave leading edge leaves the source region at $t=0.015$, whereas the wave crest detaches from the source-region boundary at $t=0.08$, when it has the highest amplitude ($\rho=1.22$). The shock formation starts at $t~\approx0.19$ and is completed at $t\approx0.32$, when it has an amplitude of $\rho=1.13$ and $v=0.13$, {\it i.e.} the formation is delayed, and the amplitude is lower compared to the case based on Equation (1). Similarly, the dip between the wavefront and the source region is somewhat shallower ($\rho=0.92$).

\section{Discussion and Conclusion} 
      \label{S-concl}

In this article we presented numerical simulations of the formation and evolution of large-amplitude MHD simple waves. We considered the very basic initial configurations to educe
general characteristics of the MHD shock formation in an idealized homogeneous environment. The main purpose is to have reference results that can be compared to the results of more sophisticated simulations that consider more realistic characteristics of the environment. Now that we have at our disposal the results of the simulations presented in this article, such a comparison will help us to distinguish which characteristics are a consequence of basic processes and which are caused by the details of a given environment.
Furthermore, it should be noted that in spite of the high level of idealization, the cylindrical configurations employed can represent, to a certain degree,
the initial stage of the coronal-wave formation by the lateral expansion of a CME during its impulsive-acceleration stage, or presumably, by the expanding leg of the flaring loop.

The most general outcome, common to all situations analyzed, is that a more impulsive source-region expansion results in a shorter time/distance needed for the shock formation, consistent with analytical considerations ({\it e.g.} \opencite{V&L00a}; \opencite{V&L00b}; \opencite{vrs01shocks}; \opencite{zic08}) and observations \cite{vrs01shocks}. The simulations show that in the most impulsive events a shock forms very close to the source-region boundary and it is initially difficult to resolve the two entities. This explains why in some studies the coronal EUV waves are (erroneously) identified as CME flanks
(for a discussion see \opencite{cheng12}). On the other hand, when the piston acceleration is low, the wave amplitude remains small and the wavefront steepening is slow. Thus, weakly accelerated eruptions are not likely to result in an observable coronal wave.

For the case of a planar magnetosonic wave we have confirmed the relationship between the wave speed and the flow speed [$w=1+3v/2$] that was analytically derived by \inlinecite{V&L00a}. At small amplitudes, the numerical simulations reproduce the Rankine--Hugoniot jump relations after the shock formation is completed. However, at large amplitudes the numerical results deviate from the analytical theory, most likely due to the numerical resolution.

From the observational point of view, the cylindrical geometry is far more interesting, since it can give us insight into the process of shock formation caused by a magnetic-arcade expansion, including the amplitude fall-off due to energy conservation \cite{zic08}. In this article we have analyzed only the most general characteristics of the perpendicular-shock formation, generated by a flux-tube expansion in an idealized homogeneous environment.
Such a process represents a two-dimensional piston mechanism of the shock-wave formation. The basic difference from the planar case (one-dimensional piston) lies in the fact that in the cylindrical geometry there are two competing effects involved in the shock-formation process. One is the nonlinear steepening of the wavefront profile (as in the planar geometry), and the other is a decrease of the wave amplitude with distance due to energy conservation (absent in planar geometry).

Of course, spherical geometry would be more relevant than the cylindrical option, since real EUV waves expand spherically. This would certainly modify the results, since the decrease rate of the wave amplitude would be governed by the $\propto r^{-2}$ effect rather than $\propto r^{-1}$. This aspect was treated in detail by \inlinecite{zic08} in a semi-analytical study which showed that the shock formation time and distance depend much more on the characteristics of the piston acceleration than on the choice of the geometry. The geometrical effect ({\it i.e.} the $\propto r^{-2}$ aspect of the energy conservation) becomes dominant only after the acceleration phase. Note that the same conclusion can be drawn from the results presented in this article by comparing the outcome for the planar and cylindrical case. In this respect, let us note that \inlinecite{zic08} have also shown that the value of the plasma-to-magnetic pressure ratio [$\beta$] does not-play a significant role as well.

In our study we considered two different types of the initial configuration: one resulting in the highest initial acceleration at the source-region boundary, and another, causing the strongest acceleration within the source-region body. The performed simulations show that, although there are differences in the evolution of the source region, the process of the shock-wave formation does not differ significantly, {\it i.e.} the evolution of the perturbation and the wave kinematics are similar in both cases.

The most important outcome of the analysis is that the formation of a perpendicular MHD shock is expected already for relatively low expansion velocities, as low as 10\,--\,20\,\% of the Alfv\'en speed. This implies that a lateral expansion of the eruption in the early phase of CMEs is a viable mechanism of the coronal wave formation \cite{ines09,patsourakos10,muhr10,veronig10,grechnev11,kozarev11,liu12,temmer12}. Furthermore, our simulations show that at the beginning of the wave formation it is difficult to distinguish the wave and the source-region expansion, especially for the case defined by Equation (2), where the strongest acceleration occurs within the source region.

The presented analysis shows that the wave initially accelerates from $w\gtrsim v_A$ to a maximum phase speed, which depends on the impulsiveness of the source-region expansion. In the decay phase, the wave-crest velocity decreases, $w\rightarrow v_A$. Thus, the initial and the late phase of the coronal wave could be used for the coronal diagnostics, since measurements of the wave kinematics in the acceleration and deceleration phase should reflect the coronal Alfv\'en speed. Similarly, the traveling density depletion that forms in the wake of the wave travels at $w\approx v_A$. Such depletions are sometimes observed in the base--difference or base--ratio EUV images, appearing as traveling coronal dimming behind the wavefront \cite{thompson00,chen02,zhukov04,muhr11}. Thus, such features can be also used to estimate $v_A$ in the quiet corona. Finally, we note that after the acceleration stage, the compression region associated with the source-region boundary becomes a stationary feature. This might be related to stationary brightenings that are sometimes observed behind the outgoing wave \cite{muhr11}.

In cylindrical geometry, the wave amplitude and the wave phase-speed are related in a relatively complex manner. In the ``driven phase", the wave amplitude at a given wave speed is higher than in the ``decay phase". In the transition between these two phases, the phase speed is almost constant for a certain period of time, while the wave amplitude decreases. This results in a loop form of the $\rho(w)$ and $v(w)$ evolutionary curves, which is consistent with observations presented by \inlinecite{muhr12}, where the dependence of the wave amplitude on the wave speed forms a closed, hysteresis-like, curve.

In Section~\ref{S-result} we have presented results in a normalized form, where velocities are expressed in units of the Alfv\'en speed and the time is expressed in units of the Alfv\'en travel time across the numerical box. As an illustration, let us assume that the diameter of the coronal source region is $2r_0=100$ Mm, which implies that the numerical box corresponds to $L=500$ Mm, since we used $r_0=0.1$. If we assume that the Alfv\'en speed in the quiet corona is in the range of $v_A=250$ km\,s$^{-1}$ \cite{WarmMann05vA}, we obtain for the Alfv\'en travel time $t_A=L/v_A=2000$ seconds. The same would be obtained for, {\it e.g.} $L=1000$ Mm and $v_A=500$ km\,s$^{-1}$.

Applying these values, one finds that the wave typically forms and steepens into a shock a few minutes after the onset of the source-region expansion. The delay is shorter for a higher source-expansion velocity, {\it i.e.} for a higher wave velocity. The distance at which the wave crest forms and detaches from the source-region boundary, {\it i.e.} the distance at which the wave should become observable, is in the range $\approx100$\,--\,200 Mm. Such time delays and starting distances are fully consistent with observations of Moreton waves ({\it e.g.} \opencite{warmuth04b}), type II bursts ({\it e.g.} \opencite{vrs95}), and EUV waves ({\it e.g.} \opencite{ines11}; \opencite{liu12}).

To conclude, the simulations presented show that already the simplest initial configurations are able to reproduce most of the features
observed in typical large-amplitude, large-scale coronal waves, including the morphology, kinematics, and scalings. Our next step will be to perform similar simulations, but employing more realistic initial configurations that depict a magnetic arcade anchored in the photosphere, and include a realistic density profile of the chromosphere, transition-region, and corona.
This will enable us to distinguish the effects that are intrinsic to the MHD wave formation from those governed by the environment.

\begin{acks}
The work presented has received funding from the European Commission's Seventh Framework Programs (FP7/2007-2013) under grant agreements No. 263252 (COMESEP project, \url{www.comesep.eu}) and No. 284461 (eHEROES project, \url{soteria-space.eu/eheroes/html/}). MT, AMV, NM, IWK acknowledge the Austrian Science Fund (FWF): FWF V195-N16 and P24092-N16. The Versatile Advection Code (VAC) was developed by G\'abor T\'oth at the Astronomical Institute at Utrecht in a collaboration with the FOM Institute for Plasma Physics, the Mathematics department at Utrecht, and the CWI at Amsterdam; in particular, Rony Keppens (FOM), Mikhail Botchev (Mathematics Dept.), and Auke van der Ploeg (CWI) contributed significantly in completing the project. G. T\'oth and R. Keppens share the responsibility and work associated with the development, maintenance, distribution, and management of the software.
We are grateful to Tayeb Aiouaz and Tibor T\"or\"ok for help in getting acquainted with VAC.
We are grateful to the referee for very constructive comments and suggestions that helped us to improve the article.
\end{acks}

\bibliographystyle{spr-mp-sola}
\bibliography{shocks}

\end{article}

\end{document}